\documentclass[%
 reprint,
 amsmath,amssymb,
 aps,
prb,
]{revtex4-2}

\usepackage{graphicx}
\usepackage{dcolumn}
\usepackage{bm}

\usepackage{siunitx}
\newcommand{\degree}{^\circ}
\newcommand{\microns}{\,\unit{\micro \meter}}

\usepackage{libertinus}

\begin{document}


\title{\sc \Large Intertwined Fano resonances in sub-wavelength metallic gratings: omnidirectional and wideband optical transmission}

\author{Denis Langevin}
 \email{denis.langevin@uca.fr}
 \affiliation{Universit\'e Clermont Auvergne, Clermont Auvergne INP, CNRS, Institut Pascal, F-63000 Clermont-Ferrand, France}
 \affiliation{
 DOTA, ONERA, Université Paris-Saclay, F-91123 Palaiseau, France
}
\author{Julien Jaeck}%
\affiliation{
 DOTA, ONERA, Université Paris-Saclay, F-91123 Palaiseau, France
}

\author{Riad Haïdar}
\affiliation{
 DOTA, ONERA, Université Paris-Saclay, F-91123 Palaiseau, France
}%
\affiliation{%
 \'Ecole polytechnique, Département de Physique, 91128 Palaiseau, France
}%
\author{Patrick Bouchon}
\affiliation{
 DOTA, ONERA, Université Paris-Saclay, F-91123 Palaiseau, France
}%

\date{\today}

\begin{abstract}
Metallic gratings can be used as infrared filters, but their performance is often limited by bandwidth restrictions due to metallic losses. 
In this work, we propose a metallic groove-slit-groove (GSG) structure that overcomes these limitations by exhibiting a large bandwidth, angularly independent, extraordinary optical transmission. Our design achieves high transmission efficiency in the longwave infrared range, driven by Fano-type resonances created through the interaction between the grooves and the central slit. This mechanism results in a tunable $2\microns$ transmission window with high rejection rate. We extend the concept to a two-dimensional GSG array, exhibiting a polarization insensitive 80\% transmission window for incident angles up to 50°, offering significant potential for infrared filtering applications.
\end{abstract}

\maketitle

Metallic films patterned with holes or slit arrays can exhibit the so-called extraordinary optical transmission phenomenon \cite{ebbesen1998extraordinary,porto1999transmission,collin2010nearly}. It was given its name since the transmission efficiency outperforms both the geometrical aperture ratio, as well as the transmission expected from diffraction theory by a single hole \cite{bethe1944theory}. 
This extraordinary transmission can be explained by the presence of plasmonic resonances, either propagative (for a periodic array of slits) or localized  in a cavity (for a Fabry-Perot type slit) \cite{garcia2010light,collin2014nanostructure}.
These seminal demonstrations have been applied to spectral filters from the visible range \cite{miyamichi2018multi} to the THz 
\cite{wu2003terahertz} and are particularly appealing for infrared multispectral imaging compared to thin film technology  \cite{haidar2010free,duempelmann2017multispectral,pelzman2018multispectral,stewart2020ultrafast,shaik2023longwave,ding2024snapshot}. 


Yet, these solutions suffer from various drawbacks. First, the  transmission follows in most cases a lorentzian shape with a full width at half maximum (FWHM) that cannot be easily modified. 
Besides, nanostructured filters that rely on propagative resonances are very sensitive to the incidence angle thus restricting them to applications with limited numerical apertures \cite{davis2017aperiodic}. 
In contrast, Fabry-Perot resonances are able to sustain angularly stable resonances, and can be obtained with a slit array in a metallic film of a typical quarter wavelength thickness \cite{astilean2000light}, but their FWHM remains determined by the material losses, thus limiting their bandwidth.

\begin{figure}[h!]
    \centering
    \includegraphics[width=0.95\linewidth]{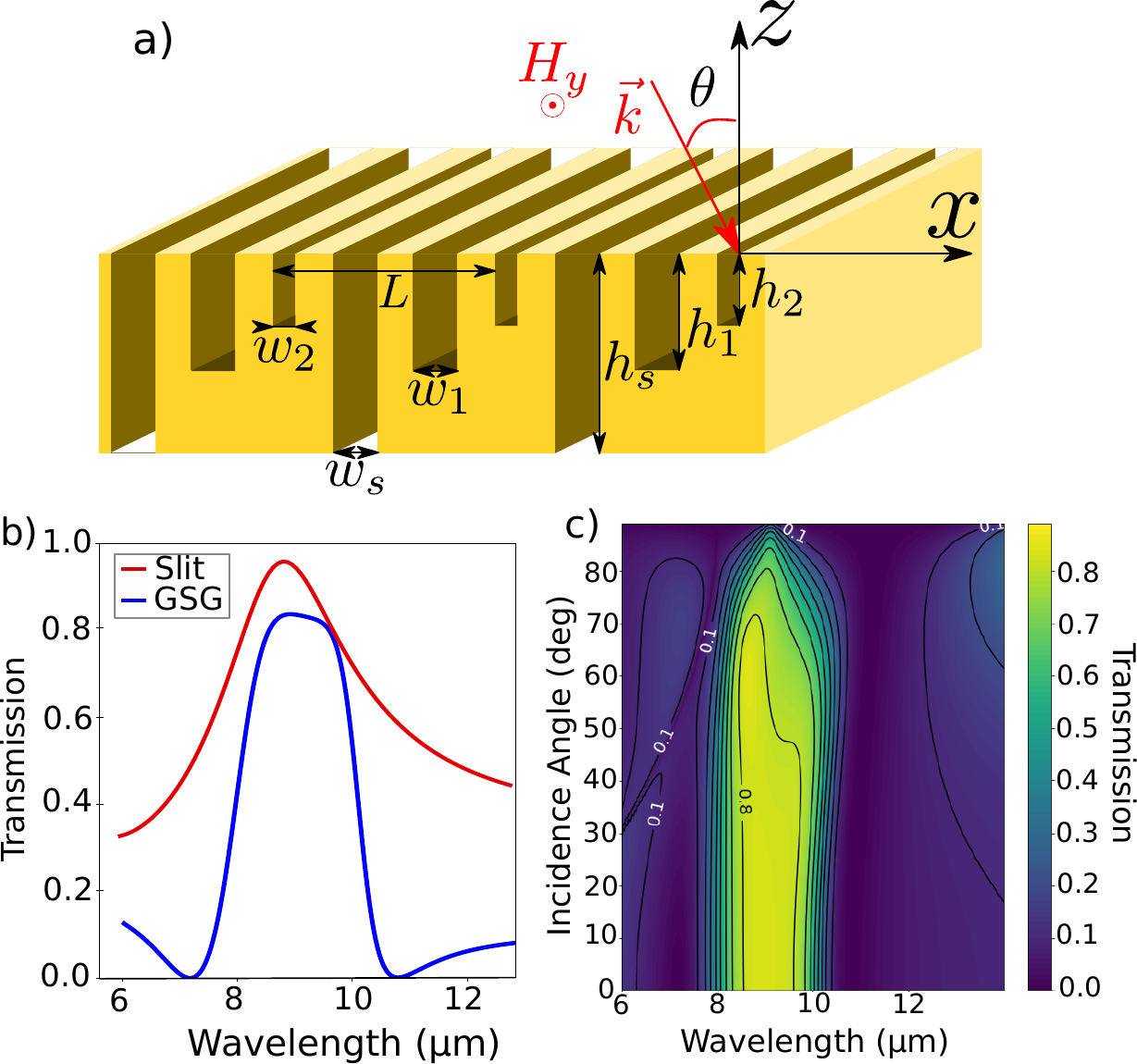}
    \caption{a) Scheme of the GSG metallic array (period $L$) and of the incoming TM-polarized light. The grooves and slit are evenly spaced throughout the period. b) Transmission ($|t^2|$) spectrum at normal incidence of the structure in gold with dimensions $w_s=w_1=0.8\microns$, $w_2=0.6\microns$, $h_s=3.6\microns$, $h_{1}=2.1\microns$, $h_{2}=1.3\microns$ and period $L=4\microns$ (blue) and of a simple slit array with same dimensions (red). c) Angular diagram ($\lambda, \theta$) of the transmission of the structure. The contour lines correspond to steps of 0.1 in the transmission.}
    \label{fig:Fig1}
\end{figure}

Here, we theoretically investigate a metallic  groove-slit-groove (GSG) structure that supports localized resonances and exhibits bandpass-like filtering in the mid-infrared. 
We begin by showing the complete structure under study in this article, and comparing it to the well-known slit-array. The rest of the article is devoted to explaining the optical behavior at play.
To this end, we show that one groove associated with a slit (SG structure), each supporting a Fabry-Perot like resonance, leads to a Fano-type profile with a larger bandwidth. 
We then show that the combination of a slit with two grooves behaves as an angularly stable transmission filter of tunable bandwidth. 
We extend these results to unpolarized filters and demonstrate an omnidirectional high transmission filter.

\begin{figure}[h!]
    \centering
    \includegraphics[width=0.95\linewidth]{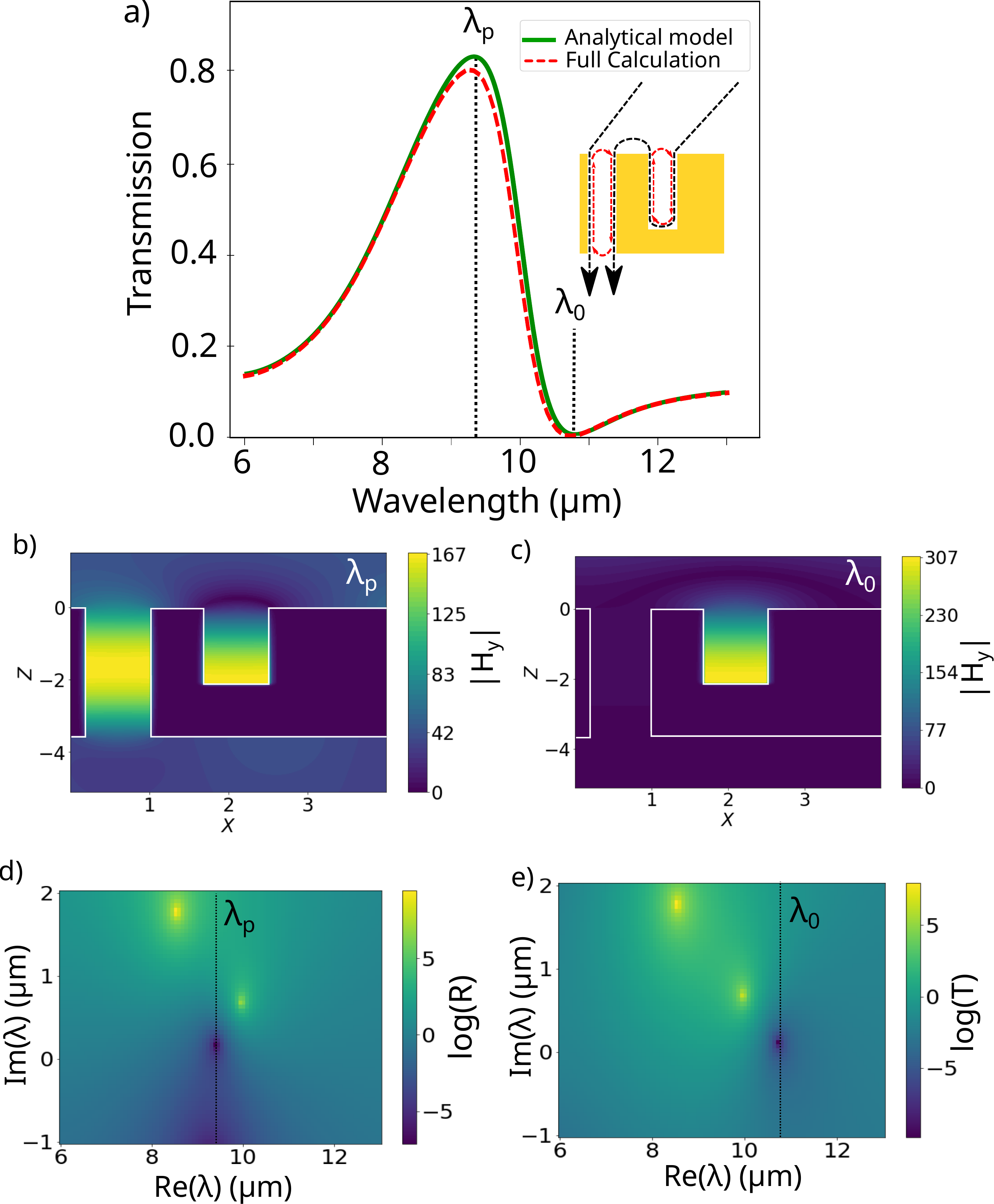}
    \caption{a) Transmission spectrum of the (SG1) slit-first groove periodic structure (same parameters), computed with an analytical model and with a full field modal method. The inset shows a scheme of the two possible interfering paths for transmitted light. b) Magnetic field amplitude $|H_y|$ distribution at the transmission peak ($\lambda_p=9.5\microns$). c) $|H_y|$ distribution at the transmission zero ($\lambda_0 =10.7\microns$). (b-c: White lines outline the structure) d) Reflexion coefficient (log scale) in the complex plane, showing the zero at $\lambda_p$. e) Transmission coefficient (log scale) in the complex plane, showing the zero at $\lambda_0$.}
    \label{fig:Fig2}
\end{figure}

The scheme of the GSG structure is presented on Fig.~\ref{fig:Fig1}a). Its unit-cell consists of one slit and two grooves repeated with a period $L$, and light is impinging on the structure with an angle $\theta$ and a transverse magnetic (TM) polarization. The metal here is gold, though it has little incidence on the overall behavior, as the wavelength is far beyond the plasma frequency.

The transmission spectrum of this GSG structure at normal incidence is shown in Fig.~\ref{fig:Fig1}b), demonstrating its band-pass behavior centered at 9$\microns$. 
The key features of this structure are first its 80\% transmission plateau with a spectral width of 1$\microns$. Second, on each side of the resonance, there is a zero of transmission and then a low value of transmission (below 10\%) that yields good rejection performance for a plasmonic filter.
To highlight the performance of our structure, the transmission spectrum of a simple slit array of same dimensions is also shown in Fig.~\ref{fig:Fig1}b). It behaves as a Fabry-Perot cavity in transmission, whose resonance wavelength is set by the optical length of the slit as $\lambda_r = 2 n_\mathrm{eff} h_s +\lambda_{\phi, s}$ (here $n_\mathrm{eff} \simeq 1.03$ due to the coupling of the surface plasmons on the cavity sidewalls and $\lambda_{\phi, s}$ accounts for the reflection phase shift at both slit interfaces) \cite{porto1999transmission}. The slits of the GSG structures also behave as Fabry-Perot cavities with a back mirror reflection, thus their resonance wavelength is given by $\lambda_r^{1,2} = 4 n_\mathrm{eff} h_{1,2} + \lambda_{\phi, g}$ \cite{bouchon2011total}.

The transmission spectrum for varying angular incidence is shown in Fig.~\ref{fig:Fig1}c), demonstrating that these filtering properties have a high angular tolerance, with the 80\% high, 2$\microns$ wide transmission bandwidth being exhibited on a $[0\degree, 50\degree]$ range.
Between $50\degree$ and $70 \degree$, the transmission remains higher than 70\% on the plateau bandwidth (8.5 to 9.5 $\microns$). The resonances spectral positions are angularly stable up to grazing incidence but the transmission amplitude and the bandwidth are gradually decreasing. Remarkably, the zero of transmission at $\lambda=10.7 \microns$ is stable up to grazing incidence, while the one at $\lambda=7.1 \microns$ is perturbed by the diffraction order for angles higher than $50 \degree$.

The angular tolerance of this structure is expected, given the nature of the resonances at play. Indeed, the Fabry-Perot cavities made by the metallic slits are known to exhibit particularly angularly stable resonances \cite{astilean2000light}. As these resonances control the optical transmission of the structure, the final transmission spectrum is also angularly stable. This also implies that the spectrum shape (zeros and maximum) are easily controlled by tuning the dimensions of the slits and grooves, as will be detailed further below.

The particular shape of the extraordinary optical transmission resonance is due to the participation of two different Fano interferences \cite{limonov2017fano,limonov2021fano}. 
Each of these Fano interferences arises from the combination of the resonances of the slit and of one of the grooves. To confirm this, the transmission of the slit and the first groove (SG1 structure) with the same dimensions and the same period is plotted in Fig.~\ref{fig:Fig2}a). It exhibits the typical lineshape of a Fano interference. The inset shows schematically the explanation of this phenomenon, \textit{i.e.} two optical paths that interfere in the structure, giving rise to interferences.
The spectra in Fig.~\ref{fig:Fig2}a) are computed independently by an analytical one-mode model \cite{Langevin2020study} and a full field modal method \cite{bouchon2010fast}. The perfect agreement between both simulations shows that only the fundamental mode has to be taken into account to explain the spectral profile, which is consistent with the Fano interference explanation.

The magnetic field amplitude maps in Fig.~\ref{fig:Fig2}b-c) show that at the resonance peak ($\lambda _p =9.5 \microns$), both the slit and the groove are concentrating the field and thus playing a role on the resonance. This hints at a coupling mechanism which slightly modifies the resonance frequencies. A detailed study of the resonant modes can be found in the Supplementary Material. On the other hand, at the resonance dip ($\lambda _p =10.7 \microns$), the field is concentrated only in the groove, prohibiting the transmission. 
To investigate further the mechanism at stake, we use the singular analysis approach in the domain of complex frequencies.
In this case, the scattering matrix can be written as $S(\omega)= A \prod_m \frac{\omega-\omega_z^{(m)}}{\omega-\omega_p^{(m)}} $, where $\omega_p^{(m)}$ and $\omega_z^{(m)}$ are poles and zeros, and $A$ a scaling constant \cite{grigoriev2013optimization}. The combination of a pole and a zero with different real part of the frequency makes a Fano profile appear, leading to the sharp slope between the maximum transmission and zero transmission \cite{grigoriev2013singular}.
The complex pole study in Fig.~\ref{fig:Fig2}d-e) shows the resonant behavior of the SG combination. The slit alone has a resonance corresponding to the pole at $Re(\lambda)=8.5\microns$. However, when combined with the slit, a pole-zero combination appears for the transmission around 10.5$\microns$, as seen on Fig.~\ref{fig:Fig2}d), leading to a sharp Fano profile.
The transmission maximum itself is better seen with the zero of the reflection on Fig.~\ref{fig:Fig2}e).

\begin{figure}[ht]
    \centering
    \includegraphics[width=0.75\linewidth]{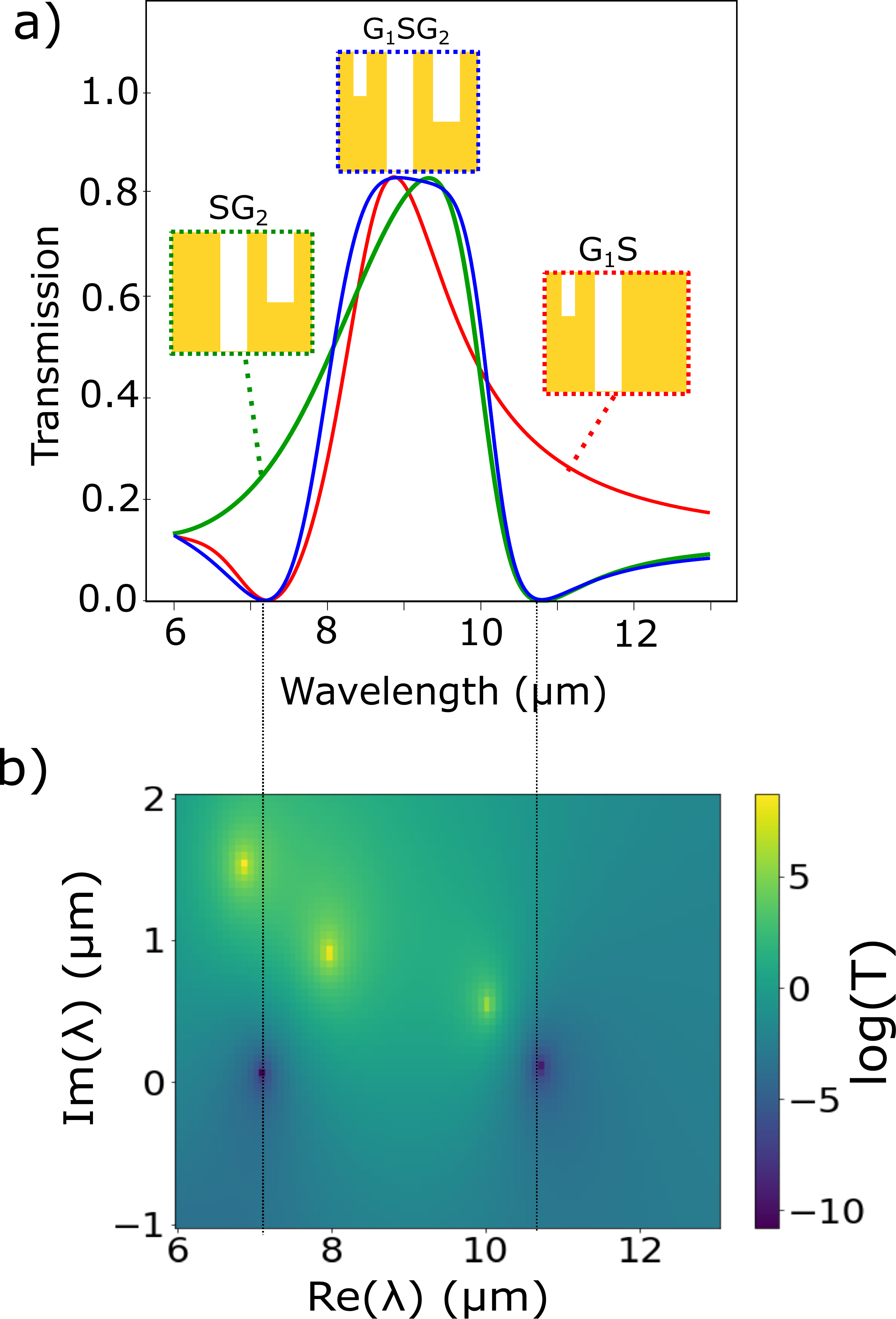}
    \caption{a) Comparison of the transmission spectra of the GSG structure (blue curve) with the $\mathrm{SG}_2$ (green curve) and the $\mathrm{G}_1\mathrm{S}$ (red curve) structures. 
    b) Transmission coefficient (in log scale) in the complex plane, showing the two zeros at $\lambda_{0}^{(1)}$ and $\lambda_{0}^{(2)}$.}
    \label{fig:Fig3}
\end{figure}

This analysis can be applied to the complete GSG structure to demonstrate that it behaves as the combination of two independent Fano profiles. 
The transmission spectrum of the GSG structure is compared to both SG structures in Fig.~\ref{fig:Fig3}.
The resonant transmission of the complete structure is surrounded by two transmission zeros at $\lambda_{0}^{(1)} = 7.25\microns$ and $\lambda_{0}^{(2)} = 10.75\microns$. Each of these zeros corresponds to a Fano profile exhibited by a given SG combination that follows the classical form $T \propto \frac{(\Omega+q)^2}{(\Omega^2+1)}$, where $q=\cot \delta$ is the Fano parameter that depends on the phase difference between the resonator and the continuum (see Supp. Mat. for values of q in SG structures) and that dictates the asymmetry of the response and $\Omega$ is the normalized frequency detuning with the resonator.
Indeed, the green (resp. red) curve on Fig.~\ref{fig:Fig3} corresponds to the transmission spectrum of the slit and the deeper groove (resp. shallower groove), exhibiting a Fano profile with zero at $\lambda_{0}^{(1)}$ (resp. $\lambda_{0}^{(2)}$).
Remarkably, when the GSG combination is done, the zeros are identical to the individual SG ones (see supplemental material).
The response of the GSG structure can be described as the product of the two individual Fano responses $T\propto \frac{(\Omega_1+q_1)^2}{(\Omega_1^2+1)}\frac{(\Omega_2+q_2)^2}{(\Omega_2^2+1)}$, and in that optimized structure, $q_1\simeq-q_2$.


The complex pole study of this full structure in Fig.~\ref{fig:Fig3}  confirms this interpretation: each zero in the transmission is close to, but not vertically aligned with, a transmission pole. The transmission band is thus widened by the combination of the two transmission peaks, and the slopes between the peak and the zeros are sharpened due to the Fano profile.

\begin{figure}[ht!]
    \centering
    \includegraphics[width=0.92\linewidth]{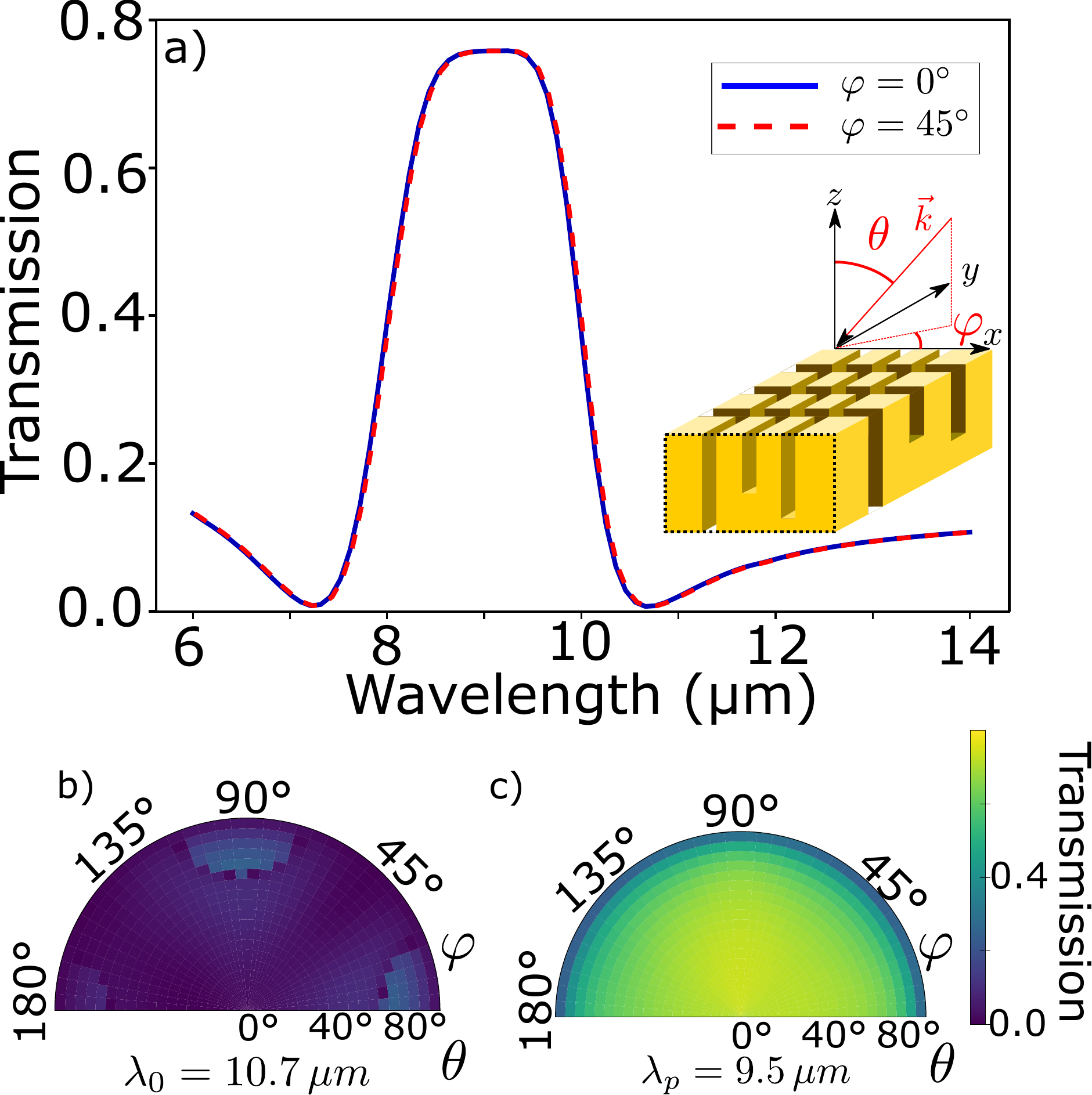}
    \caption{a) Computed transmission spectrum of the 2D bi-periodical GSG structure, for an incidence angle of $\theta = 15\degree$ and azimuthal angle of $\varphi = 0\degree$ (blue line) and $\varphi = 45\degree$ (red line). The inset shows the scheme of the 2D GSG structure's unit cell as well as the impinging light in conical mounting (angles $\theta$ and $\varphi$). The geometries of the slit and grooves are identical in the two directions. b,c) Transmission angular diagram as a function of $\theta$ and $\varphi$ at b) the transmission dip wavelength $\lambda_0 =10.7\microns$ and (c) the transmission peak wavelength $\lambda_p =9.5\microns$.}
    \label{fig:Fig4}
\end{figure}

This Fano behavior can easily be tuned by choosing different depths for the slits and grooves. Though a coupling between the slit and groove modes appears, leading to non-trivial shifts in the resonances  (see Supp. Mat.), a reliable rule of thumb for the structure design is the following.
First, the central wavelength of the transmission band is controlled by the resonance wavelength $\lambda_s \simeq 2 h_s$ of the slit that acts as a half-wavelength resonator. 
The bandwidth is controlled by the quarter-wavelength resonance of the two grooves, $\lambda_{g1} \simeq 4 h_{g1}$ and $\lambda_{g2}\simeq 4 h_{g2}$ that defined the transmission zeros on each side of the transmission band.
These values must take into account the phase offset induce by the reflections inside each cavity as wemm as the effective index \cite{collin2007waveguiding}. 
Besides, if the structure is buried or filled with a given material, all the depths must be divided by the refractive index.
Though similar "free standing" structures have already been fabricated \cite{collin2010nearly}, the use of a low refractive index substrate would slightly affect the transmission (see supp. Mat.).

In order to extend the angular stability of this filter to azimuthal variations, it is possible to use a 3D structured surface, where slits and grooves are etched in perpendicular directions. The unit cell of this structure is shown in inset of Figure~\ref{fig:Fig4}. On Figure~\ref{fig:Fig4}a) are shown transmission spectra for various incident polarization of azimuthal angle $\varphi = 0\degree$ and $45\degree$ computed with RCWA \cite{Denis_rcwa}.
These spectra show that the 3D structure has exactly the same response in transmission as the 2D structure.

Figure~\ref{fig:Fig4}b-c) shows polar representations of the evolution of the transmission at the resonance zero (Figure~\ref{fig:Fig4}b) and peak (Figure~\ref{fig:Fig4}c)) as a function of both polar and azimuthal angles. We can see that the resonance is very stable and polarization-insensitive, to the same degree as the 2D structure.

In conclusion, we have demonstrated numerically the characteristics of a band-pass filter of tunable bandwidth and wavelength, whose main asset is its large angular stability (-50°,+50°). The double Fano profile mechanism giving rise to this behavior was explained and verified, giving valuable insights to design similar structures.

\appendix
\paragraph*{Funding}
This work was supported by an AID scholarship

\paragraph*{Disclosures} The authors declare no conflicts of interest.

\paragraph*{Data availability} Data underlying the results presented in this paper can be obtained from the authors upon reasonable request.

\paragraph*{Supplemental document}
See Supplement 1 for supporting content.

\bibliography{article_EOT_open}

\end{document}


\title{\sc \Large Intertwined Fano resonances in sub-wavelength metallic gratings: omnidirectional and wideband optical transmission: supplemental document}

\begin{abstract}
This document provides supplementary information to "Intertwined Fano resonances in sub-wavelength metallic gratings: omnidirectional and wideband optical transmission". It includes the pole-zero study in the complex plane for the slit, SG and GSG structures.  Magnetic field maps are also provided at the zero and poles frequencies for the SG and GSG structures. Finally, properties of other structures, based on the same principle but designed for other wavelengths, are presented.
\end{abstract}

\maketitle



\section{Pole-zero study}

\begin{figure*}[ht]
    \centering
    \includegraphics[width=0.8\linewidth]{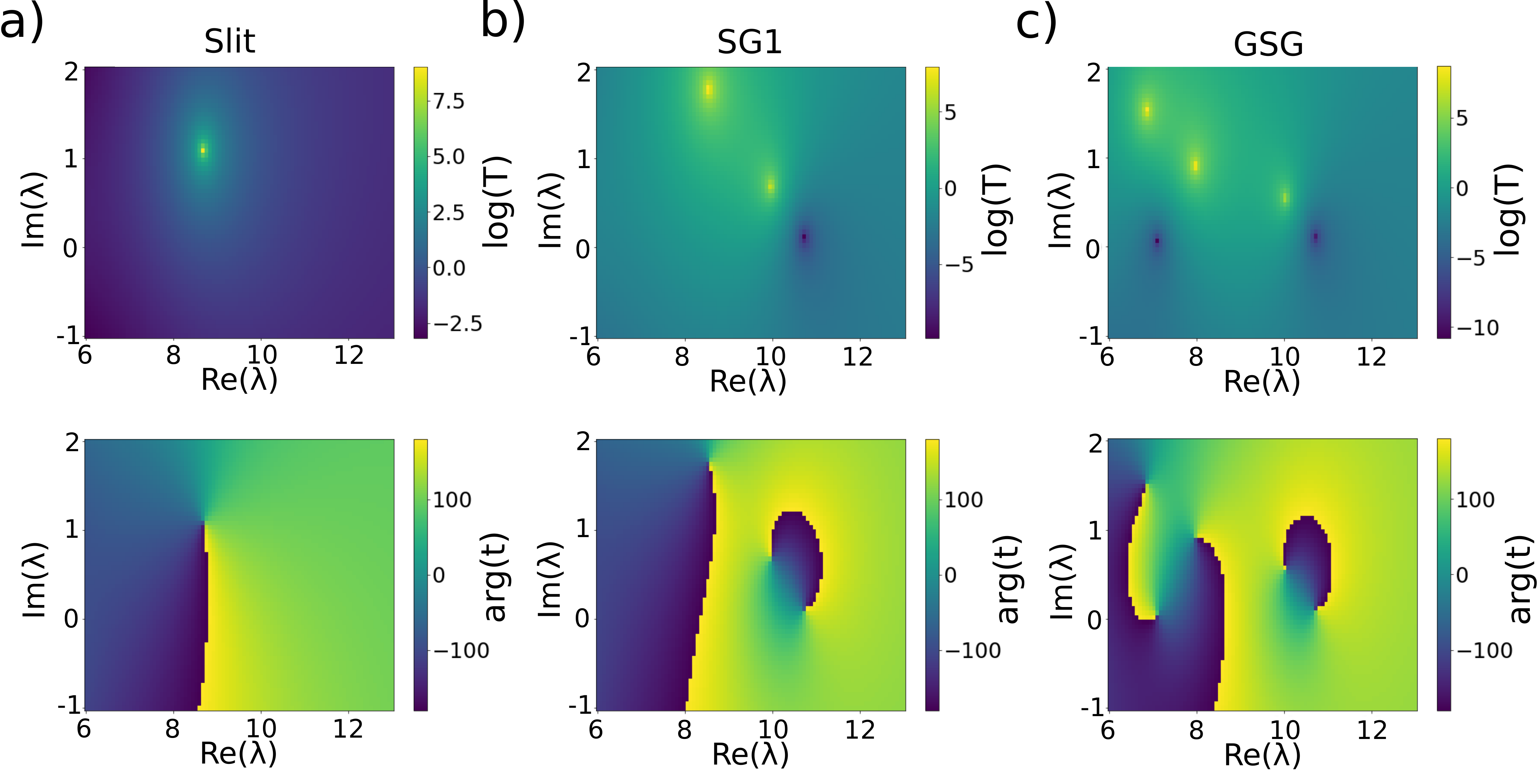}
    \caption{Amplitude and phase of the transmission in the complex plane for a) the slit, b) the SG$_1$ structure and c) the GSG structure.}
    \label{fig:FigS_full_complex}
\end{figure*}

Figure~1 shows the transmission coefficient (value and phase) as a function of a complex wavelength of incidence. This complex study gives complete information about the structure, because the poles (divergent transmission) and zeros (zero transmission) fully characterize its resonant behavior. The structure's transmission spectrum is thus a "slice" of this 2D map, along the real axis.
Figure.~1c) shows this study for the full structure, while Figure.~1b) shows it for only the first slit+groove combination and Figure.~1a) shows it for the slit alone.
The phase map of the transmission coefficient is particularly useful, as the phase jump (visible by a sudden contrast) draws a line between linked poles and zeros, and also makes visible the existence of zeros out of the computation domain.
How much the zero-phase region is tilted near the real axis is also a good prediction of how sharp the Fano resonance will be.

Comparison between these shows that the addition of a groove does not have a completely straightforward effect. Indeed, it does not simply add a pole-zero combination, otherwise the slit pole at $\lambda=8.8+1.2j\microns$ on Fig.~S1a) would not be shifted to $\lambda=8.5+1.7j\microns$ for SG1 (on Fig.~S1b)) or to $\lambda=8+j\microns$ for GSG (on Fig.~S1c)). Indeed, there is a coupling between the slit mode and the groove modes which leads to two new poles each time a groove is added.

This means that the interferential interpretation is a bit simplistic: the transmission peak (resp. dip) cannot only be explained by the fact that the groove and slit modes' reflection interfere destructively (resp. constructively) at a given wavelength.
The Fano profiles, however, are very well predicted by this complex mode study: the poles and zeros, not being symmetric with respect to the real wavelength axis, lead to quick variations of the spectrum.

\subsection{Fano parameter analysis}

The $q$ Fano parameter is the best description of the asymmetry of the spectrum (\cite{limonov2017fano}), and can be given by (\cite{grigoriev2013singular}):

\begin{equation}
    q = -\cot\left( \frac{1}{2} \arcsin\left[\frac{-2 Im(\omega_r)}{\omega_p - \omega_d}\right] \right)
\end{equation}
with $\omega_r$ the complex resonant frequency, and $\omega_{p,d}$ the real frequency of the pole/dip.

This analysis leads to:
\begin{itemize}
    \item $q\simeq-1.17$ for SG1

    \item $q\simeq1.02$ for SG2
\end{itemize}

The difference in sign corresponds to the fact that the zero appears before or after the peak. The proximity of the values account for the symmetry of the final profile.

\subsection{Field intensity maps}

\begin{figure*}[ht]
    \centering
    \includegraphics[width=0.3\linewidth]{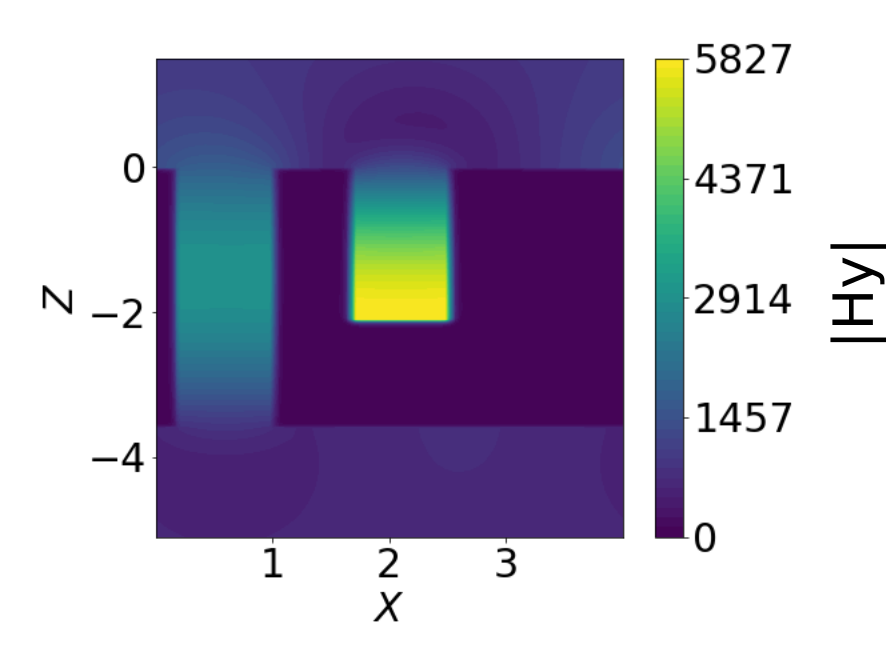}
    \includegraphics[width=0.3\linewidth]{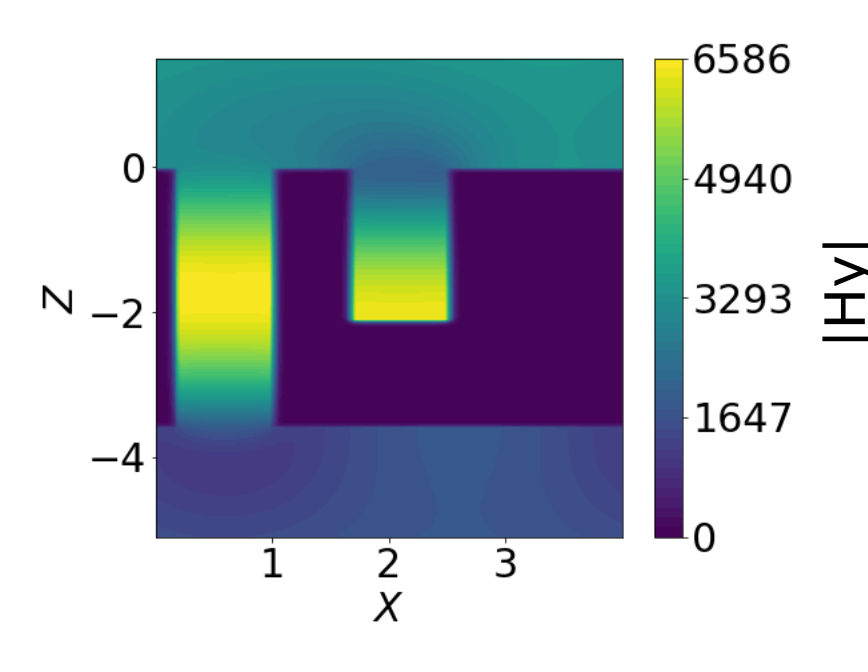}
    \caption{Magnetic field of the slit-groove (SG) structure at the two pole frequencies at $\lambda _p=10.0+0.7j\microns$ and $\lambda _p=8.5+1.8j\microns$}
    \includegraphics[width=0.3\linewidth]{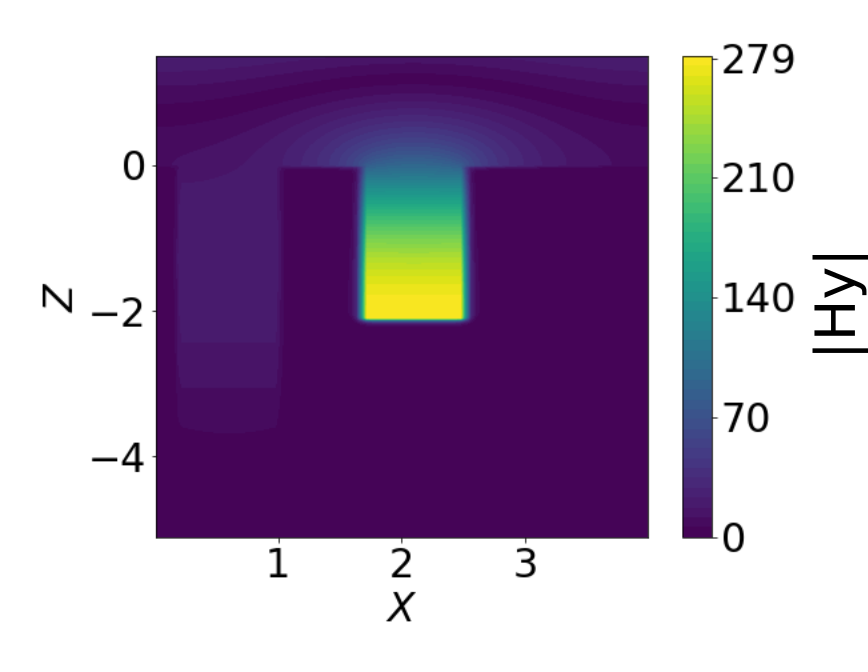}
    \caption{Magnetic field of the slit groove structure at $\lambda_0=11\microns$}
\end{figure*}

\begin{figure*}[ht]
    \centering
    \includegraphics[width=0.3\linewidth]{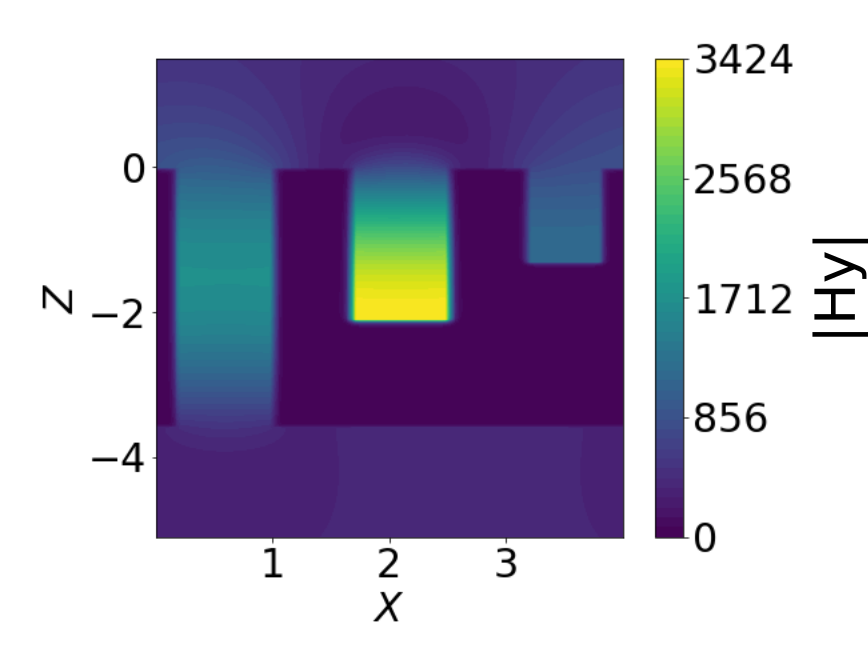}
    \includegraphics[width=0.3\linewidth]{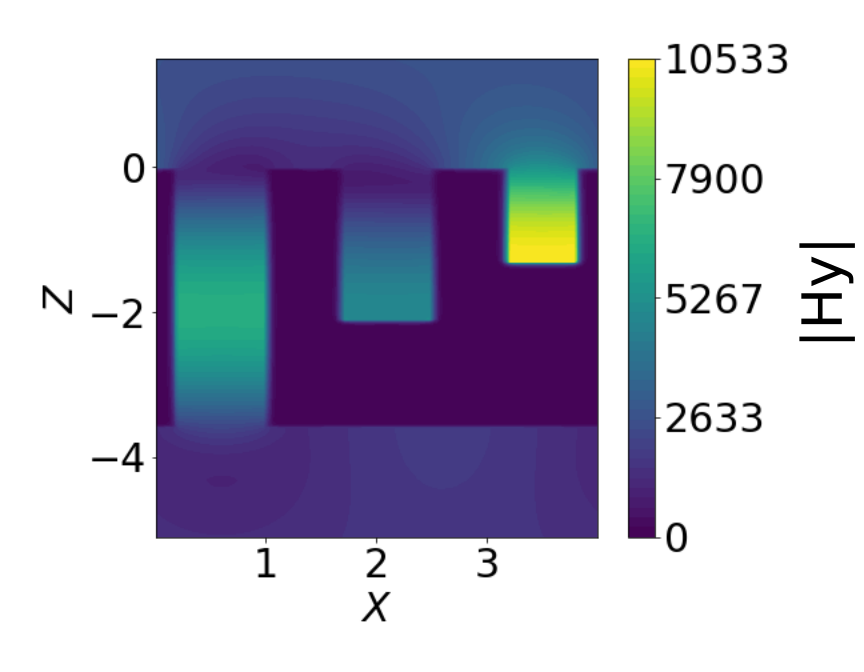}
    \includegraphics[width=0.3\linewidth]{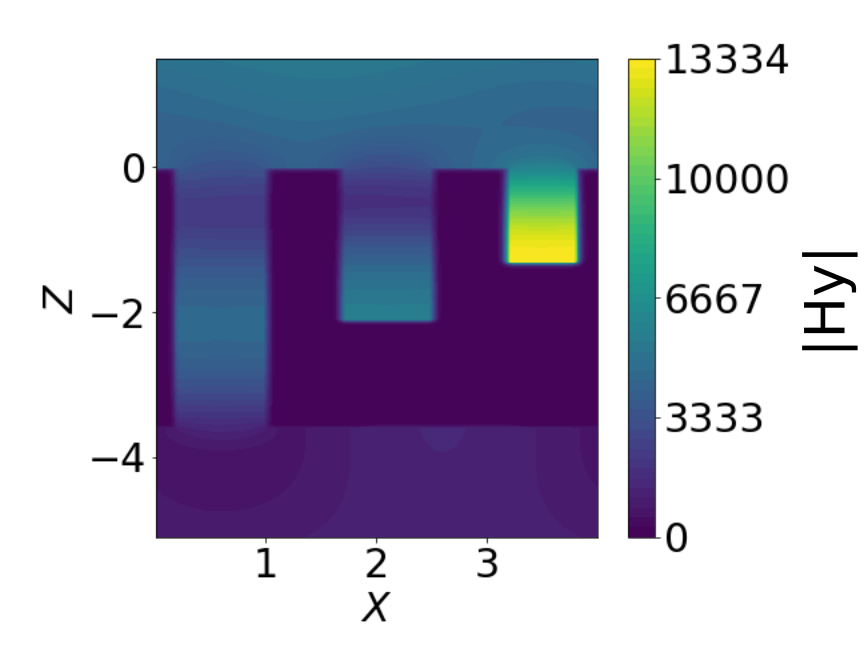}
    \caption{Magnetic field of the GSG structure at the three poles at $\lambda _p=10+0.6j\microns$, $lambda_p=8.0+0.9j\microns$ and $\lambda_p=6.8+1.5j\microns$}
    \includegraphics[width=0.3\linewidth]{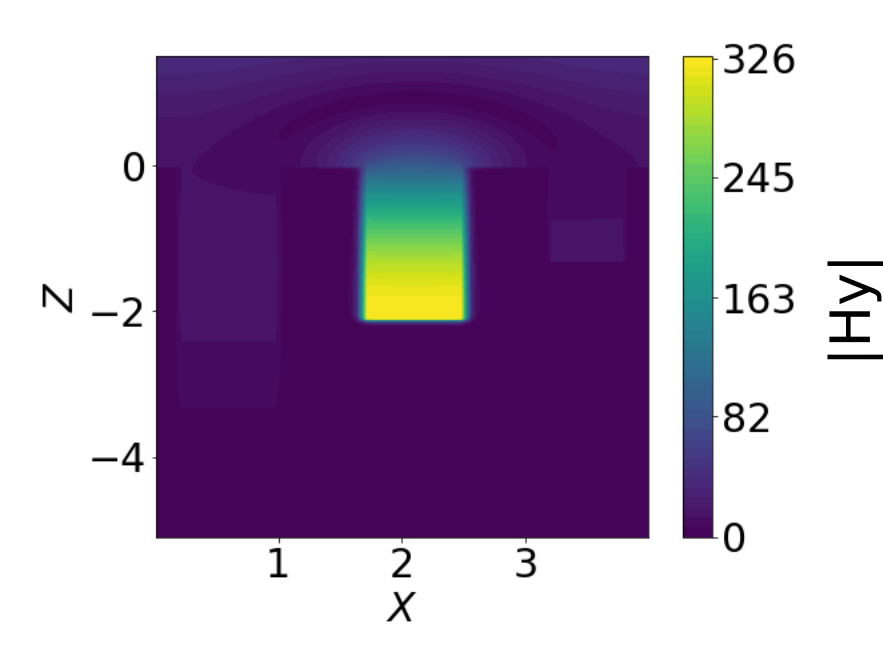}
    \includegraphics[width=0.3\linewidth]{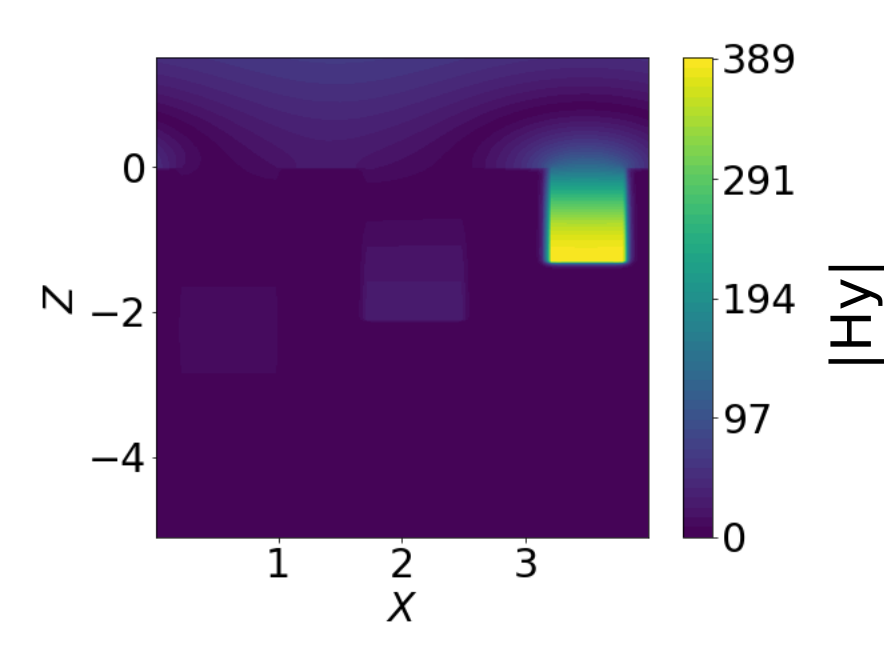}
    \caption{Magnetic field of the GSG structure at the two zeros occuring at $\lambda _0=10.6\microns$ and $\lambda _0=7.0\microns$}
\end{figure*}

Figures~2 and 3 show field intensity (precisely: $|H_y|$) maps at the poles and zeros identified in Fig.~1b) for the SG1 structure. They show clearly a very low field intensity in the slit at the transmission zeros. (at least one order of magnitude lower than at the poles).
Furthermore, a high field intensity can be observed in the grooves at most poles. This can hint at one of the device's limitation: strong resonance in a metallic groove leads to losses by absorption in the metal. This can probably explain why the transmission maximum reaches 90\% but no higher.

Figures~4 and 5 show similar field maps for the poles and zeros identified on Fig~1c), for the GSG structure.
The behavior is similar to that observed on Figs~S2 and S3 : high intensities in both a groove and the slit for poles, lower field intensities localized mostly in the grooves for zeros.

\newpage
\section{Alternative GSG structures}
\subsection{Structures for other spectral ranges}
\begin{figure*}[ht]
    \centering
    \includegraphics[width=0.28\linewidth]{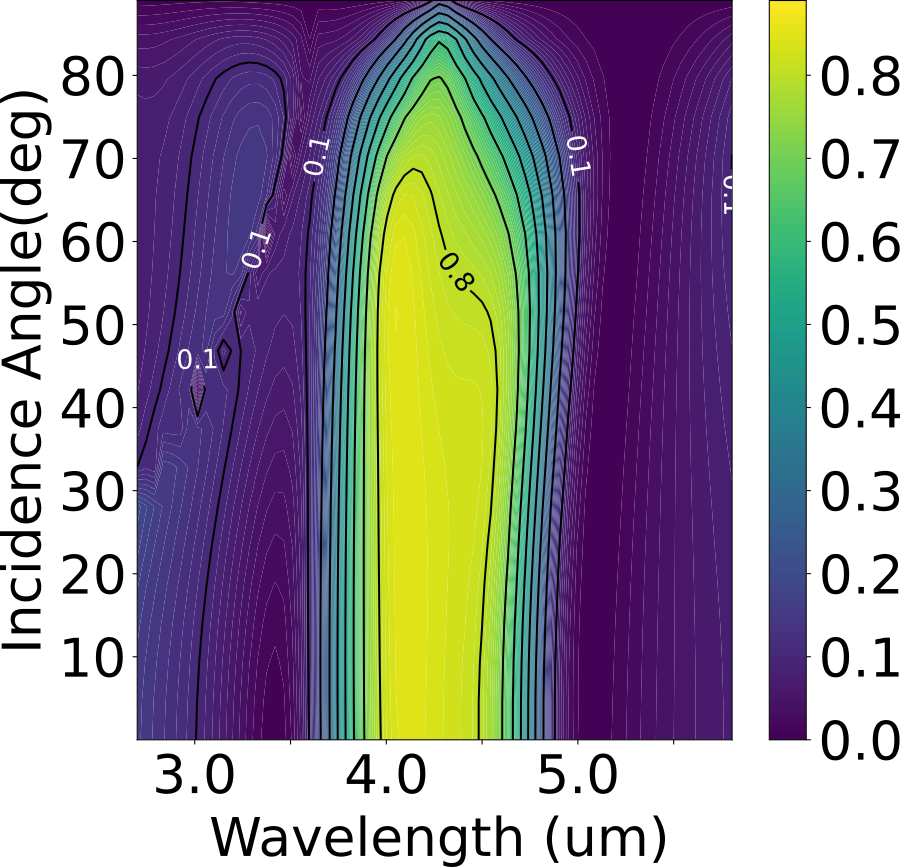}
    \quad
    \includegraphics[width=0.28\linewidth]{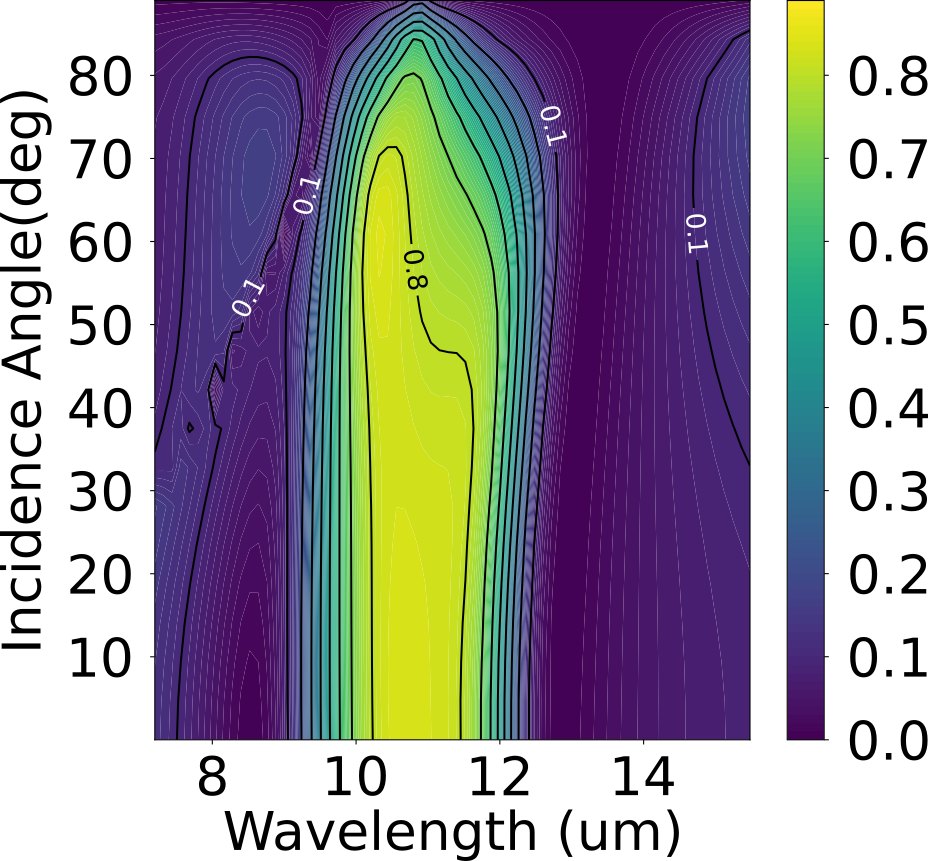}
    \quad
    \includegraphics[width=0.28\linewidth]{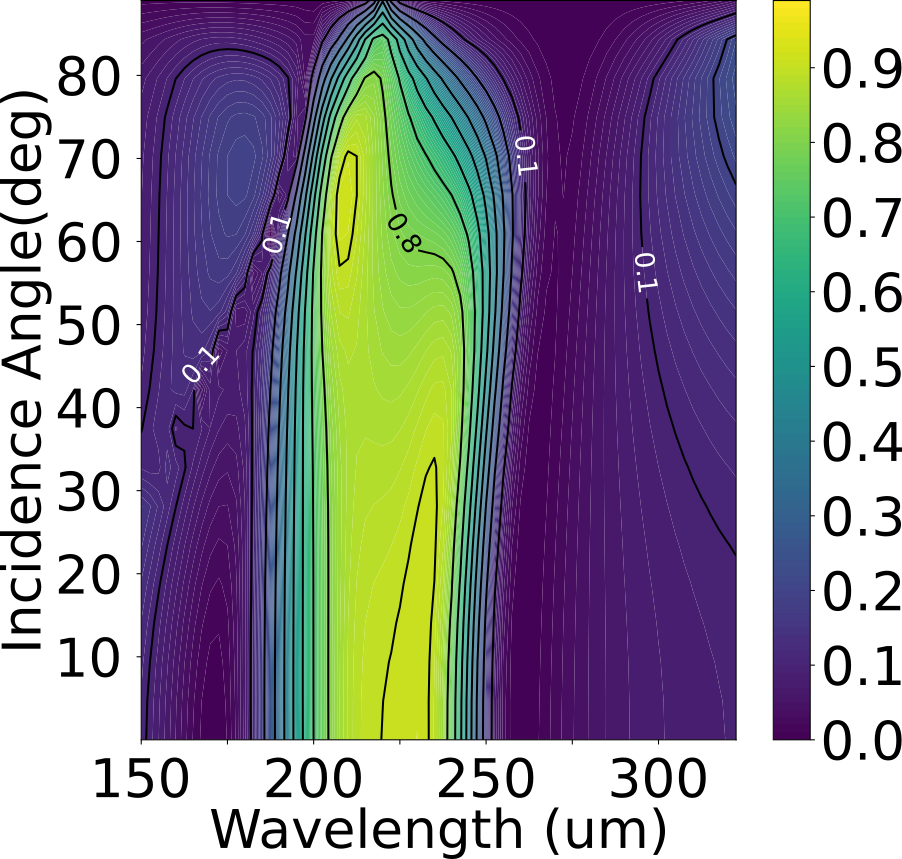}
    \caption{(left) 3-5$\microns$ structure (dimensions:  $w_s =0.36\microns$, $ w_1 = 0.36\microns$, $w_2 = 0.24\microns$, $h_s = 1.62\microns$, $h_{1} = 0.94\microns$, $h_{2} = 0.58\microns$ and period $L = 1.8\microns$ ), (center) 10-12$\microns$ structure (dimension:  $w_s =0.96\microns$, $ w_1 = 0.96\microns$, $w_2 = 0.72\microns$, $h_s = 4.32\microns$, $h_{1} = 2.52\microns$, $h_{2} = 1.56\microns$ and period $L = 4.8\microns$ ), (right) THz structure (dimensions:  $w_s =20\microns$, $ w_1 = 20\microns$, $w_2 = 15\microns$, $h_s = 90\microns$, $h_{1} = 25.5\microns$, $h_{2} = 32.5\microns$ and period $L = 100\microns$ ).}
    \label{fig:scaling}
\end{figure*}

\subsection{GSG structures on a substrate}
Figure~\ref{fig:scaling} shows the wavelength-incidence plots of structures designed for other working regimes in the infrared and terahertz range (4-5$\microns$ band, 10-12$\microns$ band, 200-250 $\microns$ band). They are obtained through a direct scaling of the main article's structure (the scaling factor is 0.45, 1.2 and 25, respectively). These results show that the scaling of the structure is a simple way to shift its spectral behavior.

\begin{figure*}[ht!]
    \centering
    \includegraphics[width=0.8\linewidth]{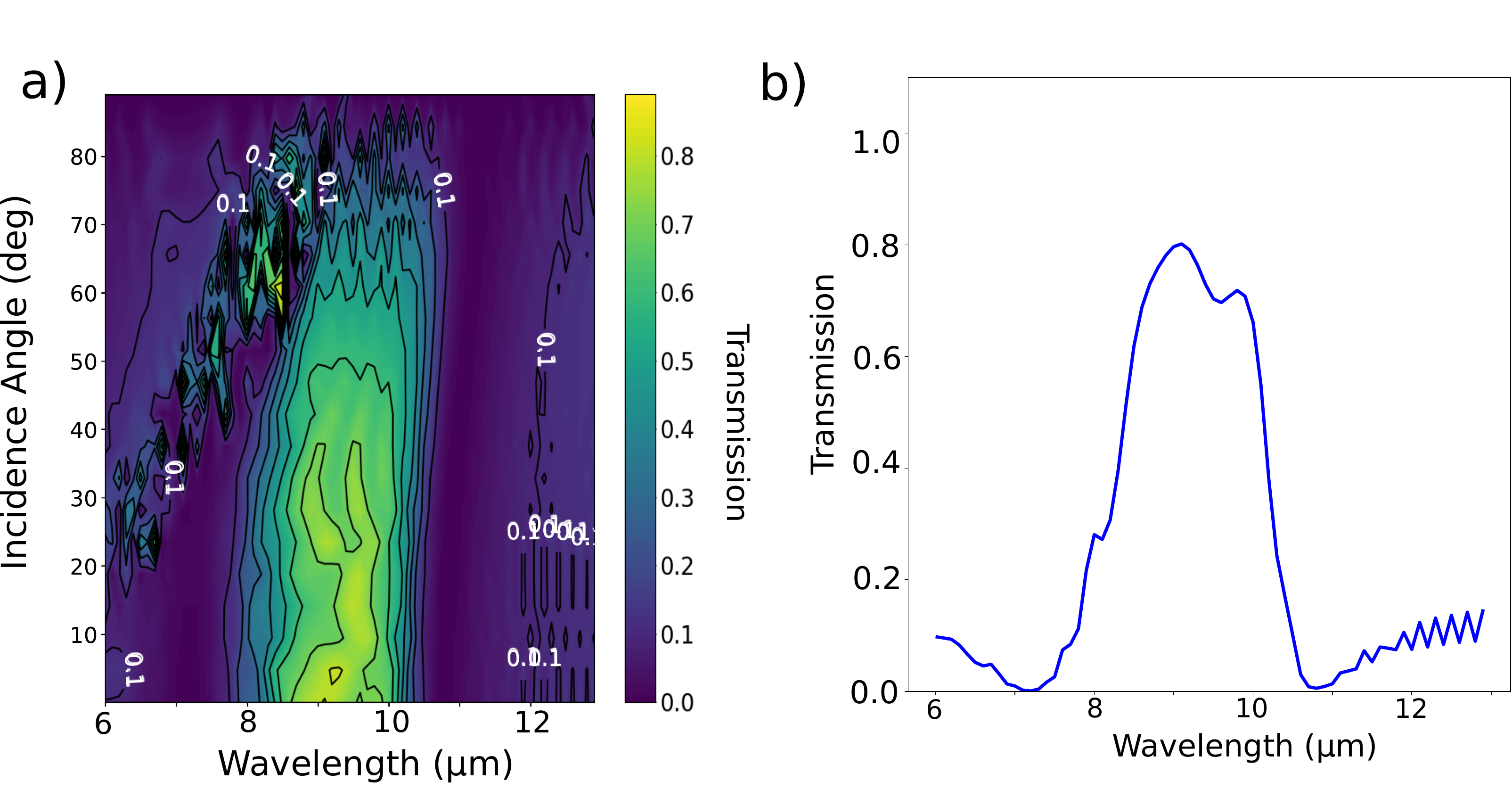}
    \caption{Structure with substrate: same dimensions as the main article's structure, with a 300$\microns$ substrate of CaF2}
    \label{fig:substrate}
\end{figure*}

Fig.~\ref{fig:substrate} shows both the wavelength-incidence plot and transmission spectrum of a structure with a 300$\microns$ thick substrate of CaF2. No antireflective coating is added. The plots show that the overall properties of the structure are conserved.


\newpage
\bibliography{article_EOT_open}